\documentclass[useAMS,usenatbib]{mn2e}

\usepackage{epsfig}

\usepackage{amsmath,amssymb,bm} 

\title[The Shape of an Accretion Disc in a Misaligned Black Hole
Binary]{The Shape of an Accretion Disc in a Misaligned Black Hole
  Binary}
  
\author[R. G. Martin, J. E. Pringle \& C. A. Tout]{Rebecca G.
  Martin,
  J. E. Pringle and Christopher A. Tout \\ University of Cambridge,
  Institute
  of Astronomy, The Observatories, Madingley Road, Cambridge
  CB3
  0HA\\}
\begin{document}

\date{}

\pagerange{\pageref{firstpage}--\pageref{lastpage}} 
\pubyear{2008}
\maketitle

\label{firstpage}

\begin{abstract}
  We model the overall shape of an accretion disc in a semi-detached
  binary system in which mass is transfered on to a spinning black
  hole the spin axis of which is misaligned with the orbital rotation
  axis.  We assume the disc is in a steady state. Its outer regions
  are subject to differential precession caused by tidal torques of
  the companion star. These tend to align the outer parts of the disc
  with the orbital plane. Its inner regions are subject to
  differential precession caused by the Lense-Thirring effect. These
  tend to align the inner parts of the disc with the spin of the black
  hole. We give full numerical solutions for the shape of the disc for
  some particular disc parameters. We then show how an analytic
  approximation to these solutions can be obtained for the case when
  the disc surface density varies as a power law with radius. These
  analytic solutions for the shape of the disc are reasonably accurate
  even for large misalignments and can be simply applied for general
  disc parameters. They are particularly useful when the numerical
  solutions would be slow.

\end{abstract}

\begin{keywords}
  accretion, accretion discs; X-rays: binaries;
\end{keywords}

\section{Introduction}

We consider a semi-detached binary system with an accretion disc
around a black hole the spin axis of which is misaligned with the
orbital plane of a companion star.  This can arise in a binary system
when one star has an asymmetric supernova explosion which gives a
velocity kick to the collapsing compact object
\citep{S70,S78,DC87,B89}.  Even a small asymmetry in the supernova can
give a large velocity kick to the newly formed black hole or neutron
star \citep{BP95,MTP09}.

The tidal torque of the companion causes each ring in the disc to
precess about an axis parallel to the binary orbital axis. The
precession rate increases with increasing radius from the black hole.
This torque can be balanced by the viscosity in the disc to create a
steady state disc if the orbital period of the companion is short
compared with the viscous timescale in the outer parts of the disc.
The disc becomes warped and twisted when the inner regions of the disc
remain aligned with the spin of the central black hole.
 
Microquasars are binary systems in which material is accreted from a
normal star on to a compact object with relativistic radio jets
\citep{MR99}. The inner accretion disc aligns with the spinning black
hole by the \cite{BP} effect. Jets produced in this region are
expected to be aligned with the spin of the black hole.  We can
measure the inclination of the jets and the inclination of the binary
orbit and find the misalignment angle.  GRO~J1655--40 might be
misaligned by an angle of at least $15^\circ$.  \cite{H95} measured a
jet inclination of $85^\circ\pm{\thinspace 2^\circ}$ to the
line-of-sight and \cite{G01} measured the orbital axis inclination to
be $70.2^\circ\pm{\thinspace 1.9^\circ}$.  For the microquasar
V4641~Sgr (SAX~J1819.3--2525), \cite{O01} deduced an orbital
inclination angle in the range $60^\circ- 70^\circ$ along the
line-of-sight and a jet inclination of less than $10^\circ$ so this
microquasar is thought to be misaligned by at least $50^\circ$.

\cite{MTP08} and \cite{MRP08} examined the microquasars GRO~J1655--40
and V4641~Sgr. They used the steady state disc model of \cite{MPT07}
warped by the \cite{BP} effect but included no binary torque term so
their discs must effectively extend to infinity in order to flatten at
the outside. Such a disc is aligned with the spinning black hole at
the centre and at the outside the inclination tends to that of the
incoming material. We consider here what effect a binary torque has on
such a steady state disc.

\section{Warped Disc Model}

We model the evolution and structure of warped accretion discs
according to the simplified formalism described by \cite{P92}.  We
consider the disc to be made up of annuli of width $dR$ and mass $2\pi
\Sigma R dR$ at radius $R$ from the central star of mass $M_1$ with
surface density $\Sigma(R,t)$ at time $t$. The angular momentum at
radius $R$ is $\bm{L}=(GM_1R)^{1/2}\Sigma \bm{l}=L\bm{l}$ where
$\bm{l}=(l_x,l_y,l_z)$ is a unit vector describing the direction of
the angular momentum of a disc annulus with $|\bm{l}|=1$.

We use equation (2.8) of \cite{P92} setting $\partial \bm{L}/ \partial
t = 0$ and add a term to describe the Lense-Thirring precession (if
the central object is a black hole) and a binary tidal torque term to
give
\begin{align}
  \frac{\partial \bm{L}}{\partial
    t}=&\frac{1}{R}\frac{\partial}{\partial R}\left[ \left(
      \frac{3R}{L} \frac{\partial}{\partial R}(\nu_1 L)
      -\frac{3}{2}\nu_1\right)\bm{L}+\frac{1}{2}\nu_2RL\frac{\partial
      \bm{l}}{\partial R}\right] \cr & + \frac{1}{R}
\frac{\partial}{\partial R} \left( \nu_2 R^2 \left| \frac{\partial
      \bm{l}}{\partial R} \right|^2 \bm{L}\right) +\bm{T}_{\rm LT}  +\bm{T}_{\rm tid}=0.
\label{maineq}
\end{align}
The Lense-Thirring precession term is given by
\begin{equation}
\bm{T}_{\rm LT}=   \bm{\Omega}_{\rm LT} \times \bm{L}= \frac{\bm{\omega}_{\rm LT}
    \times \bm{L}}{R^3},
\end{equation}
where
\begin{equation}
\bm{\omega}_{\rm LT} =\frac{2G\bm{J}}{c^2}
\label{omegap}
\end{equation}
\citep{KP85}. The angular momentum of the black hole $\bm{J}=J\bm{j}$,
with $\bm{j}=(j_x,j_y,j_z)$ and $|\bm{j}|=1$, can be expressed in
terms of the dimensionless spin parameter $a$ such that
\begin{equation}
J=acM_1\left(\frac{GM_1}{c^2}\right).
\label{angmom}
\end{equation}
This approximation of the torque is valid if $\Omega_{\rm LT} \ll a
\Omega$ where $\Omega^2=GM/R^3$ \citep{KP85}. This corresponds to $R
\gg R_{\rm g} = GM/c^2$.  The inner edge of an accretion disc is the
innermost stable orbit around the black hole. For a non-rotating black
hole this is at $6\, R_{\rm g}$ and decreases down to $R_{\rm g}$ for
a maximally rotating black hole.  The binary torque term is given by
\begin{equation}
\bm{T}_{\rm tid}=\bm{\Omega}_{\rm tid} \times \bm{L}.
\end{equation}
We discuss this torque further in the next Section.

There are two viscosities. First $\nu_1$ corresponds to the azimuthal
shear (the viscosity normally associated with accretion discs) and
secondly $\nu_2$ corresponds to the vertical shear in the disc which
smoothes out the twist. The second viscosity acts when the disc is
non-planar.  With the $\alpha$-parameterisation the viscosities are
given by $\nu_1=\alpha_1 c_{\rm s} H$ and $\nu_2=\alpha_2 c_{\rm s} H$
where $c_{\rm s}$ is the sound speed and $H$ is the scale height of
the disc. The dimensionless parameters $\alpha_1 \le 1$ and $\alpha_2$
must be determined experimentally.   \cite{PP83} find that the
viscosities can be related by
\begin{equation}
\frac{\nu_2}{\nu_1}=\frac{\alpha_2}{\alpha_1}=\frac{1}{2\alpha_1^2}
\label{alpha}
\end{equation}
to first order.  

The more detailed analysis of warped discs by \cite{Ogilvie99} and
\cite{Ogilvie00} supports this approach. They include, into
equation~(\ref{maineq}), an extra precession term of the form
\begin{equation}
\frac{1}{R}\frac{\partial}{\partial R}\left( \nu_3 R L \bm{l}\times \frac{\partial \bm{l}}{\partial R}\right),
\end{equation}
where $\nu_3$ is a third viscosity parameter related to the
precessional effects with $\nu_3=\alpha_3 c_{\rm s} H$. However,
\cite{LP07} find that $\alpha_3 \ll \alpha_2$ and we may neglect this
term.

In order to obtain analytic results we take both viscosities to have a
power law form so that
\begin{equation}
\nu_1=\nu_{10}\left(\frac{R}{R_0}\right)^\beta ~~~ {\rm and} ~~~ 
\nu_2=\nu_{20}\left(\frac{R}{R_0}\right)^\beta ,
\label{viscs}
\end{equation}
where $\nu_{10}$, $\nu_{20}$ and $\beta$ are all constants and $R_0$
is some fixed radius. In a steady flat (unwarped) disc the surface
density far from the inner edge is then
\begin{equation}
\Sigma= \Sigma_0 \left(\frac{R}{R_0}\right)^{-\beta+x},
\end{equation}
where $x=0$ for a steady accretion disc.  For a steady decretion disc
in which matter is added at a steady rate at the inner radius $x =
1/2$.  This value of $x$ gives the radial dependence of the surface
density at late times in the evolution of a decretion disc of fixed
mass. An accretion disc reaches steady state on the viscous timescale
given by
\begin{equation}
\tau_{\nu_1}=\frac{R^2}{\nu_1}.
\end{equation}
As long as the mass transfer rate remains constant over this timescale
then the disc can reach this steady state solution.

\section{Tidal Torque}

\cite{L00} and \cite{OD01} find the tidal torque in the frame of the
binary orbit, of radius $R_{\rm b}$, to be approximately
\begin{equation}
  \mathbf{T}_{\rm tid}=-\frac{GM_2R\Sigma }{2R_{\rm b}^2}
  \left[b_{3/2}^{(1)}\left(\frac{R}{R_{\rm b}}\right)\right]
  (\bm{e_z}.\bm{l})(\bm{e_z}\times \bm{l})
\end{equation}
to first order in the angle of tilt between the disc and the orbital
plane. Here $\bm{e_z}$ is a unit vector in the direction of the binary
orbital axis, $M_2$ is the mass of the binary companion star and
$R_{\rm b}$ is the binary separation.  The Laplace coefficient of
celestial mechanics can be approximated for small $z$ by
\begin{equation}
  b_{3/2}^{(1)} \left(z\right) = 
  \frac{3}{2}z \left[ 1+ \frac{15}{8}z^2+\frac{175}{64}z^4+...\right].
\label{laplace}
\end{equation}
We note that the tidal torque satisfies $\bm{l}.\bm{T}_{\rm tid}=0$ so
that the tidal torque is zero when the disc and orbit are aligned.
Thus even for large tilt angles the tidal torque term is likely to
have this form, although with modified coefficients. This tidal torque
is averaged over the orbital period of the system which we assume to
be much shorter than the viscous timescale at the outer edge of the
disc.

\section{Relevant disc radii}

We expect the centre of the disc to be aligned with the spin of the
central black hole by the Bardeen-Petterson effect. We expect the
tidal torques to align the outer regions with the binary orbit by an
analogous physical process.  We note that if the outer parts of the
disc are closer to counter-alignment than alignment, then the whole
disc is counter-rotating. The inner parts of the disc are flat and
counter-aligned with the black hole spin. The counter aligned disc
solution is oppositely twisted to an aligned solution
\citep{SF,MPT07}.

The torque on the black hole is always towards alignment of the black
hole with the outer parts of the disc because the angular momentum of
the disc is effectively very large compared to that of the hole
\citep{KLOP}. This is because the angular momentum of the binary orbit
is transferred to the outer parts of the disc when the tidal torque
acts \citep{MTP08}.  

By comparing the sizes of various terms in the governing
equation~(\ref{maineq}) we are able to get an initial idea of the
structure that the disc is likely to have when it is in a steady
state. In either case, aligned or counter aligned discs, there are
three radii of physical significance.

\subsection{The Lense-Thirring radius $R_{\rm LT}$}

A Lense-Thirring radius $R_{\rm LT}$ where the effects of
Lense-Thirring precession and viscosity balance was defined by
\cite{SF} and \cite{MPT07}
\begin{equation}
  R_{\rm LT}=\left(\frac{2 \omega_{\rm LT}}
    {\nu_{20}R_0^{-\beta}}\right)^{1/(1+\beta)}
\end{equation}
by balancing the first term of equation~(\ref{maineq}) involving
$\nu_2$ with the Lense-Thirring term and replacing $\partial/\partial
R$ with $1/R$. Well inside this radius we expect the disc to be flat
and aligned with the spin of the central black hole.

\subsection{The tidal radius $R_{\rm tid}$}

We define in a similar way a tidal radius, $R_{\rm tid}$, to be where
the tidal term balances the same viscous term in the disc so that
\begin{equation}
  R_{\rm tid}=\left(\frac{2 \nu_{20}R_0^{-\beta}(GM_{\rm
        1})^{\frac{1}{2}}R_{\rm b}^3}
    {3 GM_2 }\right)^{2/(7-2\beta)}.
\label{btw}
\end{equation}
If a disc extends beyond this radius then the viscous effects are
dominated by the binary torque term. Thus we expect that well outside
this radius the binary torque term, coupled with the viscosity,
flattens the disc and aligns with the binary orbital plane. The
precise form of $R_{\rm tid}$ will become clearer in Section~\ref{62}
when we use it to simplify equation~(\ref{eq}).

\subsection{The warp radius $R_{\rm warp}$}

The overall warping of the disc is caused by the misalignment between
the Lense-Thirring precession which dominates at small radii and the
tidal precession which dominates at large radii. The rate of warping
of the disc with radius is likely to be greatest when these two
precessional effects balance. We call this the
warp radius. It is
\begin{equation}
  R_{\rm warp}= \left(\frac{4 \omega_{\rm p} M_1^{\frac{1}{2}}R_{\rm b}^3}
    {3 G^{\frac{1}{2}} M_2}\right)^{2/9}.
\end{equation}
It is independent of the magnitude of the viscosity because we are now
balancing just the two external torques.

We may then write
\begin{equation}
  R_{\rm warp}= \left(R_{\rm LT}^{1+\beta}
    R_{\rm tid}^{(7 - 2 \beta)/2} \right)^{2/9}.
\label{bal}
\end{equation}
Thus over all we expect the disc to be warped at radii between the
Lense-Thirring radius $R_{\rm LT}$ and the tidal radius $R_{\rm tid}$,
with the rate of warping greatest at around the warp radius
$R_{\rm warp}$.

\section{Numerical Models}
\label{sec:num}

We are unable to solve the full non-linear disc evolution equation
analytically and thus in general it is necessary to solve the
equations numerically. We use the numerical scheme described by
\cite{P92}, including both a binary torque precession term and a
Lense-Thirring precession term. The binary torque is implemented in
the form
\begin{equation}
  \bm{\Omega}_{\rm tid}= \frac{1}{\tau_{\nu_2}(R_{\rm tid}) }
  \left(\frac{R}{R_{\rm tid}}\right)^{3/2}   
  \bm{e_z},
\end{equation}
with $\tau_{\nu_2}=R^2/\nu_2$, corresponding to just the first term in
the expansion in equation~(\ref{laplace}). We have assumed that the
warping is small so that $\bm{e}_z.\bm{l}\approx 1$ \citep{OD01}. The
Lense-Thirring term is
\begin{equation}
\bm{\Omega}_{\rm LT} =\frac{1}{\tau_{\nu_2}(R_{\rm LT}) }
  \left(\frac{R}{R_{\rm LT}}\right)^{-3} \bm{j} .
\end{equation}
We use 200 logarithmically distributed grid points between the inner
radius $R_{\rm in}$ and the outer radius $R_{\rm out} = 10^4 \,R_{\rm
  in}$.

We choose two different set of boundary conditions. In the first we
have zero torque at the boundaries so that $\Sigma=0$ at $R=R_{\rm
  in}$ and $R=R_{\rm out}$ which allows mass to flow freely through
both boundaries. The surface density is chosen initially to correspond
to a steady state for a flat accretion disc with the corresponding
accretion rate $\dot M$,
\begin{align}
  \Sigma & =  \begin{cases}
\displaystyle {\,\,\, \frac{\dot M}{3 \pi \nu_1}\left[1-\left(\frac{R}{R_{\rm
          in}}\right)^\frac{1}{2}\right]} & \text{if  $R<R_{\rm add}$}, \cr \cr
\displaystyle{\,\,\, \frac{\dot M}{3 \pi \nu_1}\left[\left(\frac{R_{\rm
          add}}{R}\right)^\frac{1}{2}-\left(\frac{R_{\rm
          in}}{R}\right)^\frac{1}{2}\right]} & \text{if $R>R_{\rm add}$}
\end{cases}
\end{align}
\citep{P92}. We add mass at $R_{\rm add} = 9000\, R_{\rm in}$ at a
rate that keeps the surface density constant at that point with
angular momentum corresponding to the local Keplerian values and
aligned with $\bm{e}_z$. We expect this to be in the outer flat part
of the disc.  Inside this radii the disc acts as a standard accretion
disc. We set $\partial \bm{l}/\partial R=0$ at $R=R_{\rm in}$ and
$R=R_{\rm out}$.

These boundary conditions are realistic but to compare directly with
our analytic models (Section~\ref{an}) we use a second set of boundary
conditions. We compute numerical models in which initially
$\Sigma=\dot M/(3\pi\nu_1)$ and the surface density is fixed at
$R_{\rm in}$ and $R_{\rm out}$. The disc is aligned with the black
hole at the inside so at $R=R_{\rm in}$, $\bm{l}=(\sin \theta_0,0,\cos
\theta_0)$ and at the outside $R=R_{\rm out}$, $\bm{l}=(0,0,1)$.

We choose the viscosities $\nu_1 = 0.1\, (R/R_{\rm in})^2$ and $\nu_2
= 1.0 \,(R/R_{\rm in})^2$ so that $\beta = 2$ and the local viscous
timescale $\tau_\nu = R^2/\nu$ is independent of radius. This
minimises computational run times. We have chosen $\nu_1/\nu_2=0.1$
which corresponds to $\alpha_1=0.22$ with equation~(\ref{alpha}).
With SPH simulations of warped discs, \cite{LP07} find that the ratio
$\nu_1/\nu_2$ is small.  Because $\Omega_{\rm LT} \propto R^{-3}$ and
$\Omega_{\rm tid} \propto R^{3/2} $ we may fix the various relevant
radii by choosing the strengths of the tidal and Lense-Thirring terms.
Models shown in the next sections are at a time of $500 \,( R_{\rm
  in}^3/GM_1)^{\frac{1}{2}}=50\, \tau_{\nu_1}=500\,\tau_{\nu_2}$ so
that they have reached steady state.

\subsection{Misalignment of $\theta_0=20^\circ$}

\begin{figure*}
  \epsfxsize=8.4cm
  \epsfbox{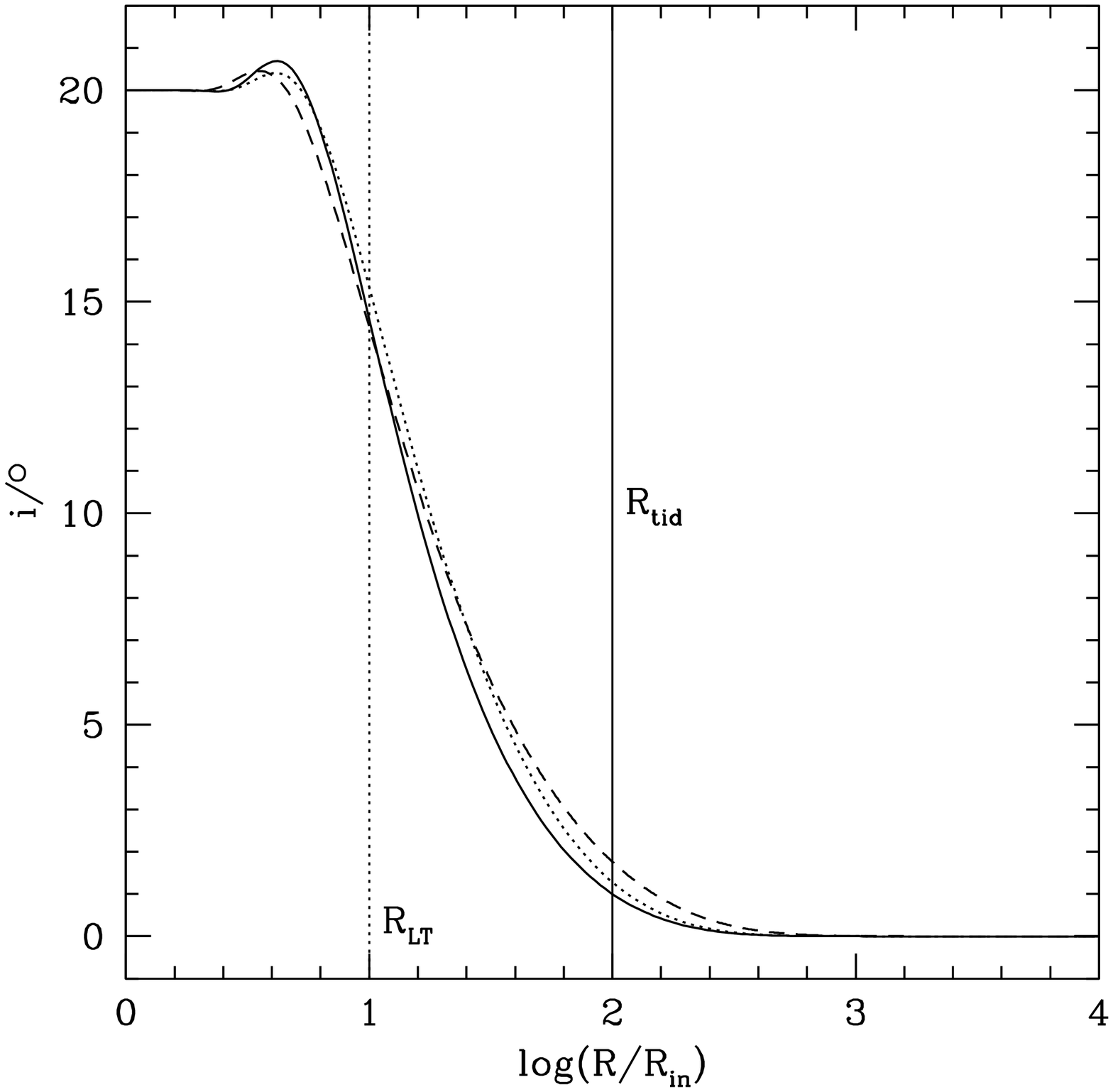}
 \epsfxsize=8.4cm \epsfbox{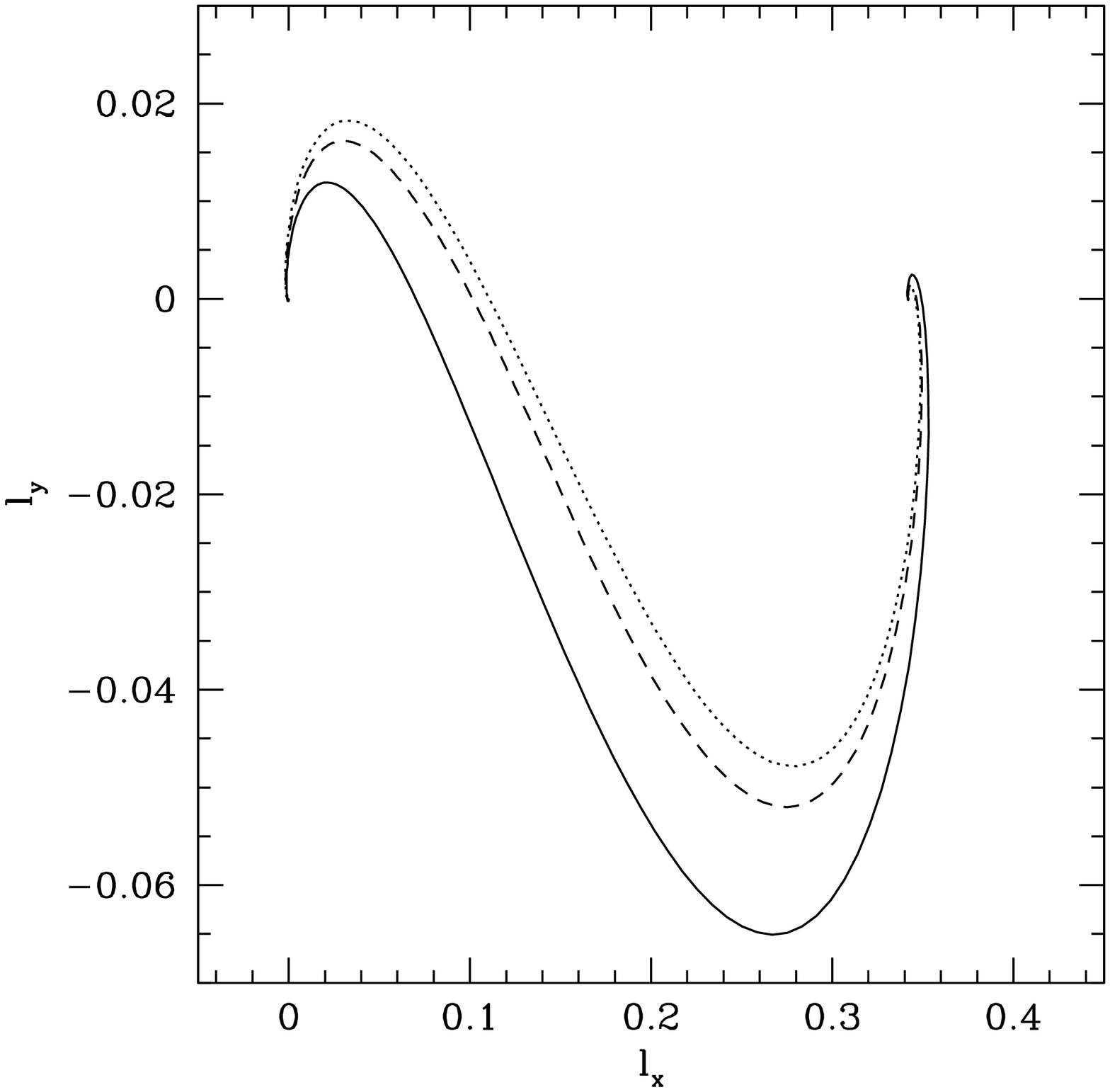}
  \caption[] {Models with the black hole spin misaligned 
    to the binary orbital axis by $\theta_0=20^\circ$.  Left: The
    inclination of the disc. Right: The twist of the disc.  The models
    are of an accretion disc with both the Lense-Thirring effect and a
    binary torque term with $R_{\rm LT}=10\,R_{\rm in}$, $R_{\rm
      tid}=10^2\,R_{\rm in}$ and $R_{\rm out}=10^4\,R_{\rm in}$.  The
    solid lines have the boundary conditions $\Sigma=0$ at $R_{\rm
      in}$ and $R_{\rm out}$.  The dotted lines are a numerical model
    with $\Sigma=\dot M/(3\pi\nu_1)$ at $R_{\rm in}$ and $R_{\rm out}$
    which corresponds to the analytic boundary conditions. The dashed
    line is the matched analytical solution described in
    Section~\ref{sec:match}.  All models have $\nu_1, \nu_2 \propto
    R^2$.  }
\label{num}
\end{figure*}

We first choose parameters so the Lense-Thirring radius is $R_{\rm LT}
= 10\,R_{\rm in}$ and the tidal radius is $R_{\rm tid} = 100\,R_{\rm
  in}$. We have $10\, R_{\rm g}<R_{\rm LT} <60 \, R_{\rm g}$ and so we
can use the approximation to the Bardeen-Petterson effect in
equation~(\ref{omegap}).  For $\beta=2$ the warp radius is then
$R_{\rm warp} = 21.5\,R_{\rm in}$ (equation~\ref{bal}).  The outer
disc and tidal precession is aligned with the $z$--axis. The
Lense-Thirring precession axis is tilted at an angle of $\theta_0 =
20^\circ$ to the $z$--axis. The tilt angle of the disc as a function
of radius is shown in the left panel of Fig.~\ref{num}.  The relevant
radii are also marked.  Note that the disc itself is not only tilted
but that the tilt varies with radius so that the overall shape of the
disc is a twisted warp (right panel).

\begin{figure}
  \epsfxsize=8.4cm
  \epsfbox{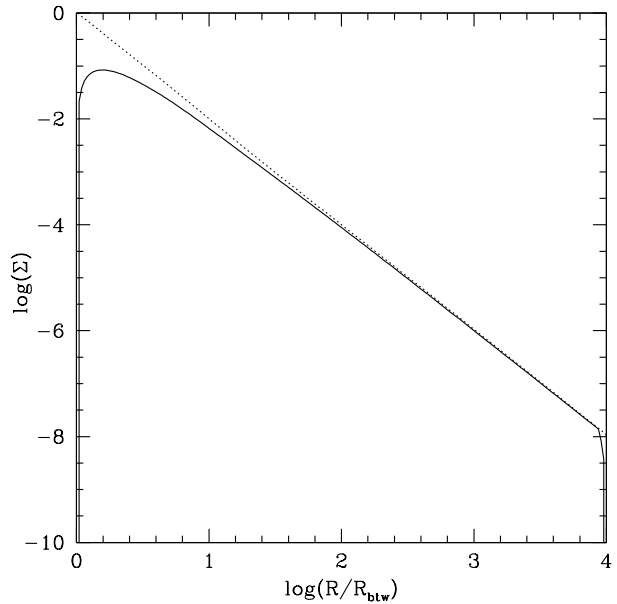}
  \caption[] {  The surface density for numerical models with black hole spin
    misaligned to the binary orbital axis by $\theta_0=20^\circ$.  An
    accretion disc with both the Lense-Thirring effect and a binary
    torque term with $R_{\rm LT}=10\,R_{\rm in}$, $R_{\rm
      tid}=10^2\,R_{\rm in}$ and $R_{\rm out}=10^4\,R_{\rm in}$.  All
    models have $\nu_1, \nu_2 \propto R^2$.  The solid line has the
    zero torque boundary conditions $\Sigma=0$ at $R_{\rm in}$ and
    $R_{\rm out}$.  The dotted line has $\Sigma=\dot M/(3\pi\nu_1)$ at
    $R_{\rm in}$ and $R_{\rm out}$ as in our analytic models.}
\label{sig}
\end{figure}

We see that the maximum rate of change of tilt does occur at about
$R_{\rm warp}$. Once we are well within $R_{\rm LT}$ or well outside
$R_{\rm tid}$ the disc does indeed become locally flat and aligned
with the locally dominant precession axis. We see that both the
numerical models with different boundary conditions and the analytical
model are very similar for this small misalignment angle.  In
Fig.~\ref{sig} we plot the surface density for the two numerical
models in Fig.~\ref{num}. The models are similar except in the regions
close to the boundaries as expected.

\subsection{Changing the Warp Radii}

Here we demonstrate the effect of changing the outer two warp radii.
We have three models each with the constant inclination and surface
density boundary conditions. We choose $R_{\rm LT}$ and $R_{\rm tid}$
so that $R_{\rm warp} = 21.5$ remains unchanged.  The first model has
the same warp radii as that in Fig.~\ref{num} with $R_{\rm LT} =
10\,R_{\rm in}$ and $R_{\rm tid} = 100\,R_{\rm in}$. The second has
$R_{\rm LT} = 3.16\,R_{\rm in}$ and $R_{\rm tid} = 10^3\,R_{\rm in}$
and the third $R_{\rm LT} = R_{\rm tid}= R_{\rm warp}=21.5\,R_{\rm
  in}$. We use $\Sigma=\dot M/(3\pi\nu_1)$ at $R_{\rm in}$ and $R_{\rm
  out}$.

\begin{figure*}
 \epsfxsize=8.4cm \epsfbox{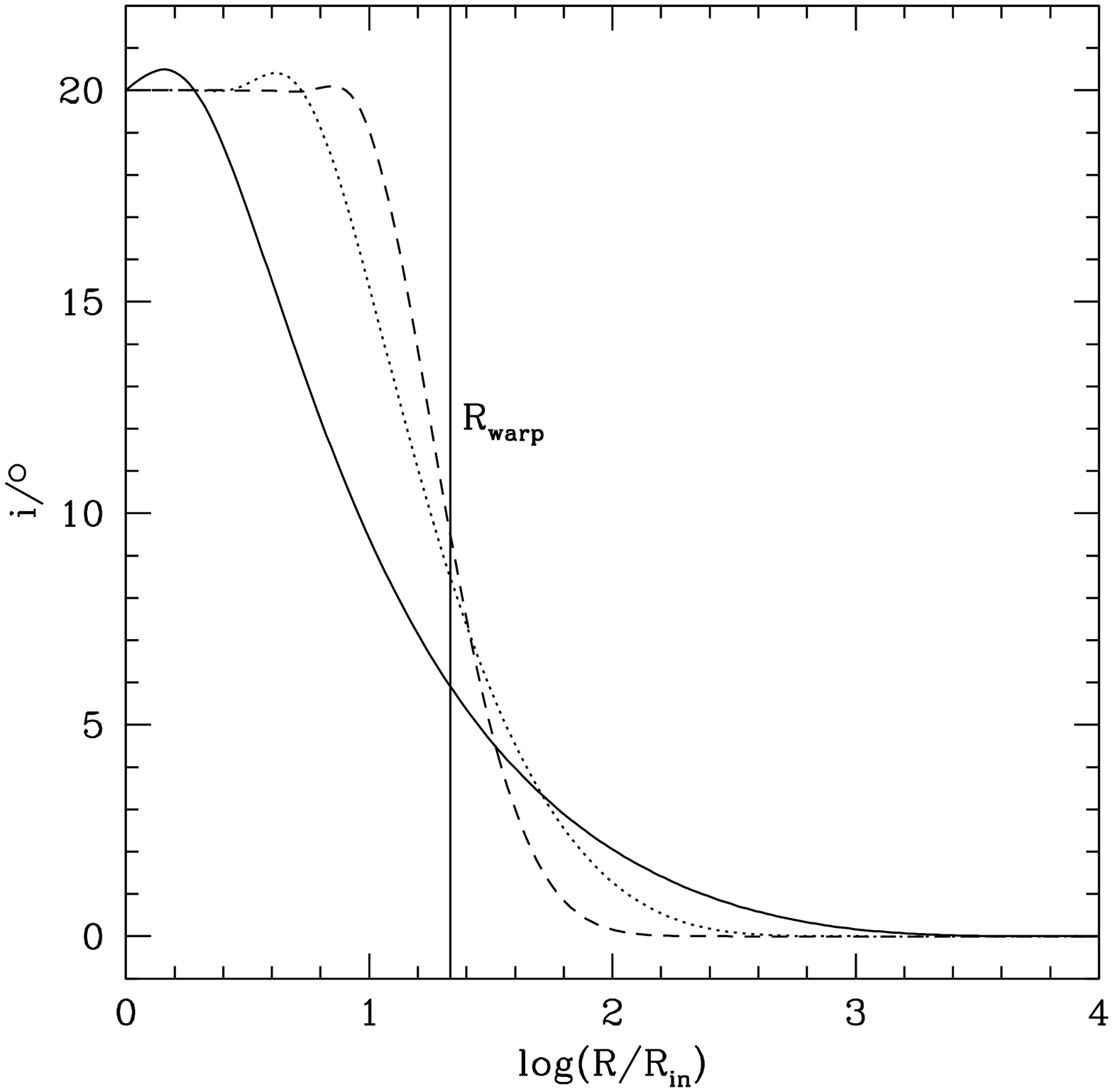}
 \epsfxsize=8.4cm \epsfbox{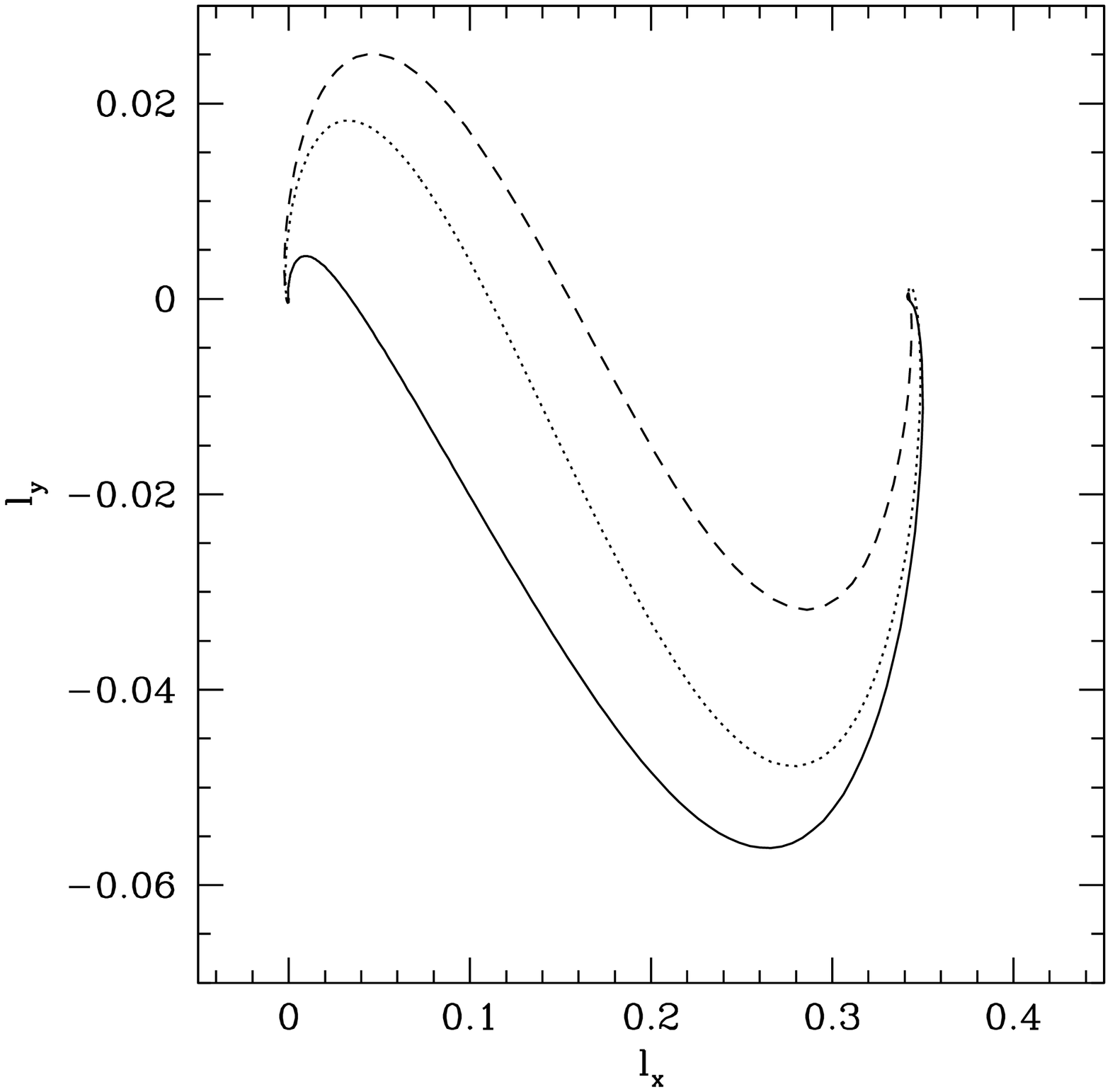}
  \caption[] {Numerical models of an accretion disc with 
    both the Lense-Thirring effect and a binary torque term with the
    constant inclination and surface density boundary conditions. The
    black hole spin is misaligned to the binary orbital axis by
    $20^\circ$ and $\nu_1$ and $\nu_2 \propto R^2$. All three models
    have $R_{\rm warp}=21.5\, R_{\rm in}$. The solid lines have
    $R_{\rm LT}=3.16\,R_{\rm in}$ and $R_{\rm tid}=10^3\,R_{\rm in}$.
    The dotted lines are the numerical models in Fig.~\ref{num} with
    $R_{\rm LT}=10\, R_{\rm in}$ and $R_{\rm tid}=100\,R_{\rm in}$.
    The dashed lines have $R_{\rm warp}=R_{\rm LT}=R_{\rm tid}=21.5
    \,R_{\rm in}$. We use $\Sigma=\dot M/(3\pi\nu_1)$ at $R_{\rm in}$
    and $R_{\rm out}$.}
\label{num2}
\end{figure*}

In Fig.~\ref{num2} we plot the inclinations and twists of these
models. The steepest rate of change of tilt angle occurs at around
$R_{\rm warp}$ in all models. As $R_{\rm tid}$ increases the outer
parts of the disc become aligned with the outer precession vector
(orbital plane) at larger radii. As $R_{\rm LT}$ decreases the inner
disc aligns with the black hole spin at smaller radii.  So $R_{\rm
  warp}$ tells us where the disc is warped $R_{\rm LT}$ and $R_{\rm
  tid}$ tell us how sharp the warp is.

\subsection{Misalignment of $\theta_0=60^\circ$}

\begin{figure*}
 \epsfxsize=8.4cm \epsfbox{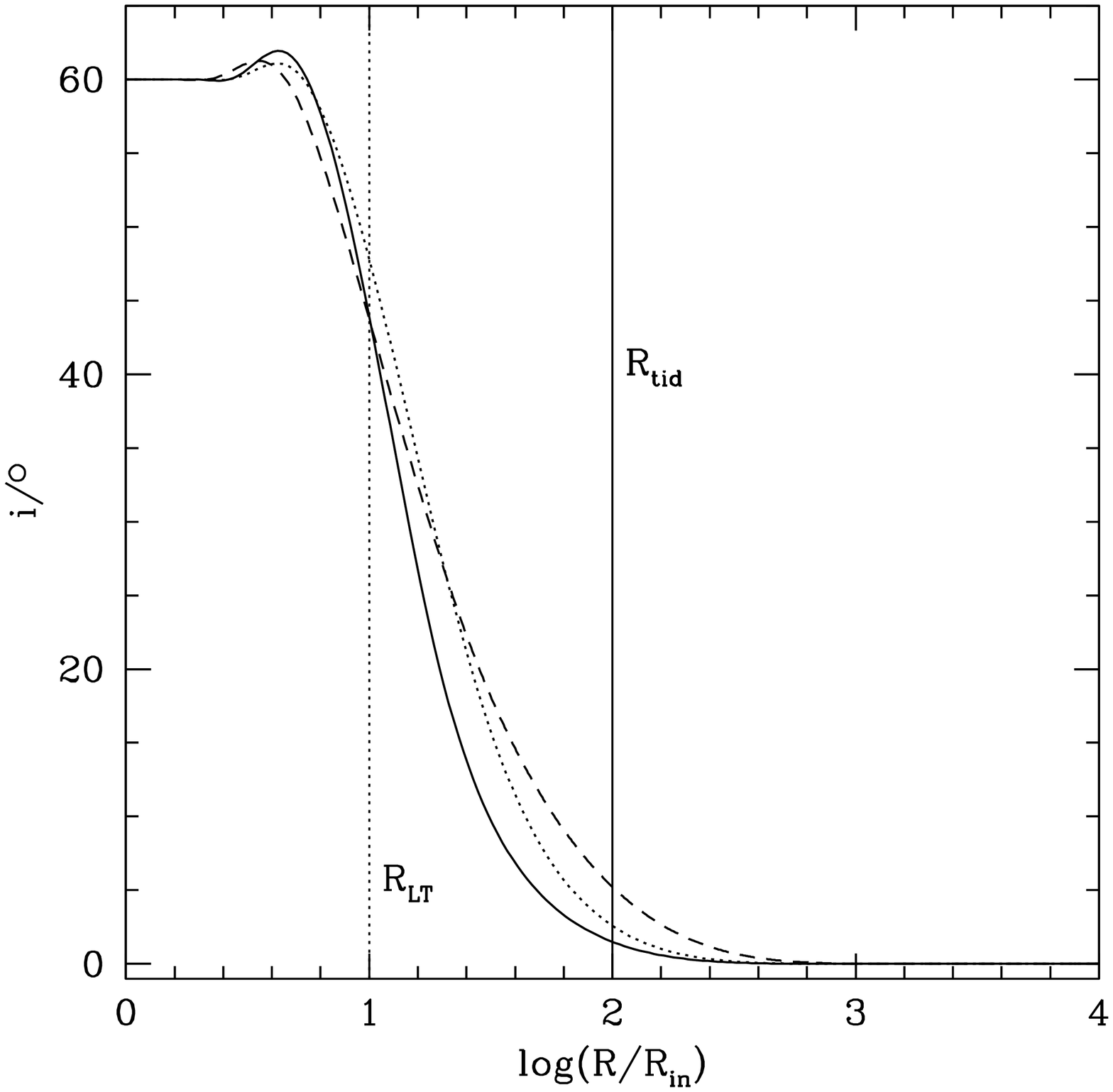}
 \epsfxsize=8.4cm \epsfbox{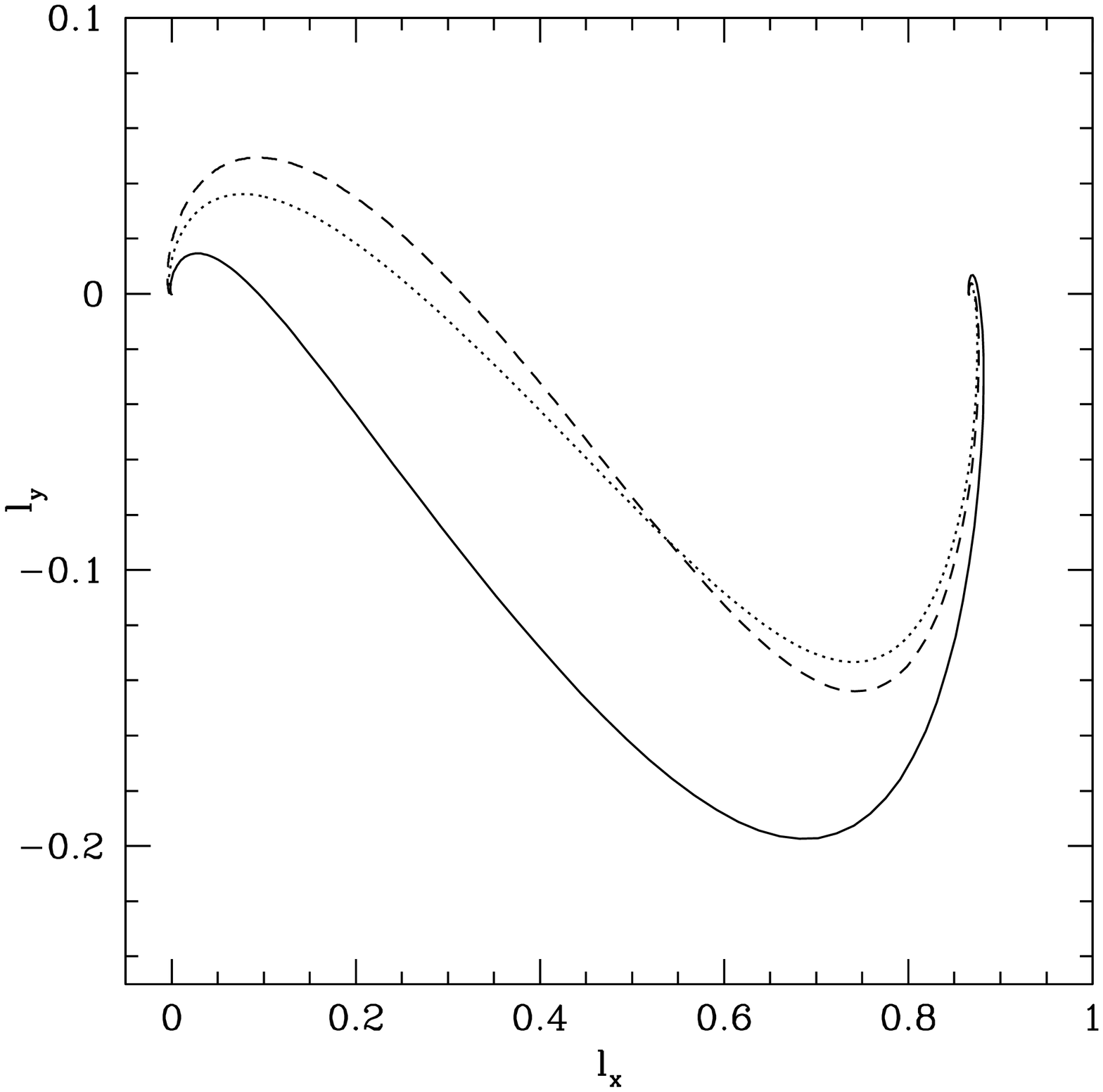}
  \caption[] { Models with the black hole spin misaligned
    to the binary orbital axis by $\theta_0=60^\circ$.  The models
    show an accretion disc with both the Lense-Thirring effect and a
    binary torque term with $R_{\rm LT}=10\,R_{\rm in}$, $R_{\rm
      tid}=10^2\,R_{\rm in}$ and $R_{\rm out}=10^4\,R_{\rm in}$. The
    solid lines have zero torque boundary conditions. The dotted lines
    are numerical models with $\Sigma=\dot M/(3\pi\nu_1)$ at $R_{\rm
      in}$ and $R_{\rm out}$. This corresponds to the analytic
    boundary conditions. The dashed line is an equivalent analytical
    matched model.  All models have $\nu_1, \nu_2 \propto R^2$.}
\label{num3}
\end{figure*}

In Fig.~\ref{num3} we plot the inclination and twist of a disc which
is misaligned by $60^\circ$. The numerical model with the fixed
inclination and surface density boundary conditions is similar to the
analytic solution even at this high inclination.

\subsection{Misalignment of $\theta_0=80^\circ$}

\begin{figure*}
 \epsfxsize=8.4cm \epsfbox{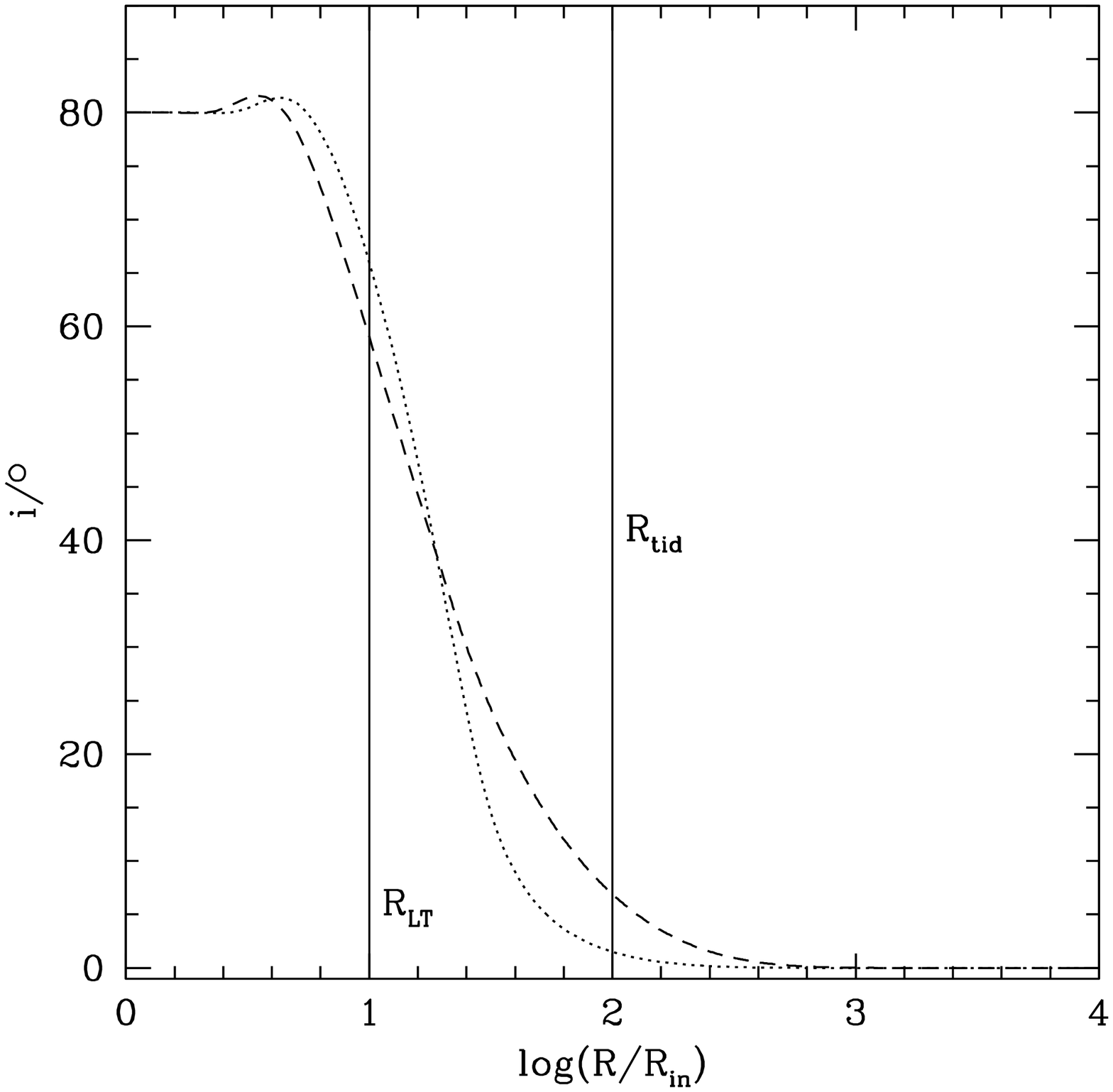}
 \epsfxsize=8.4cm \epsfbox{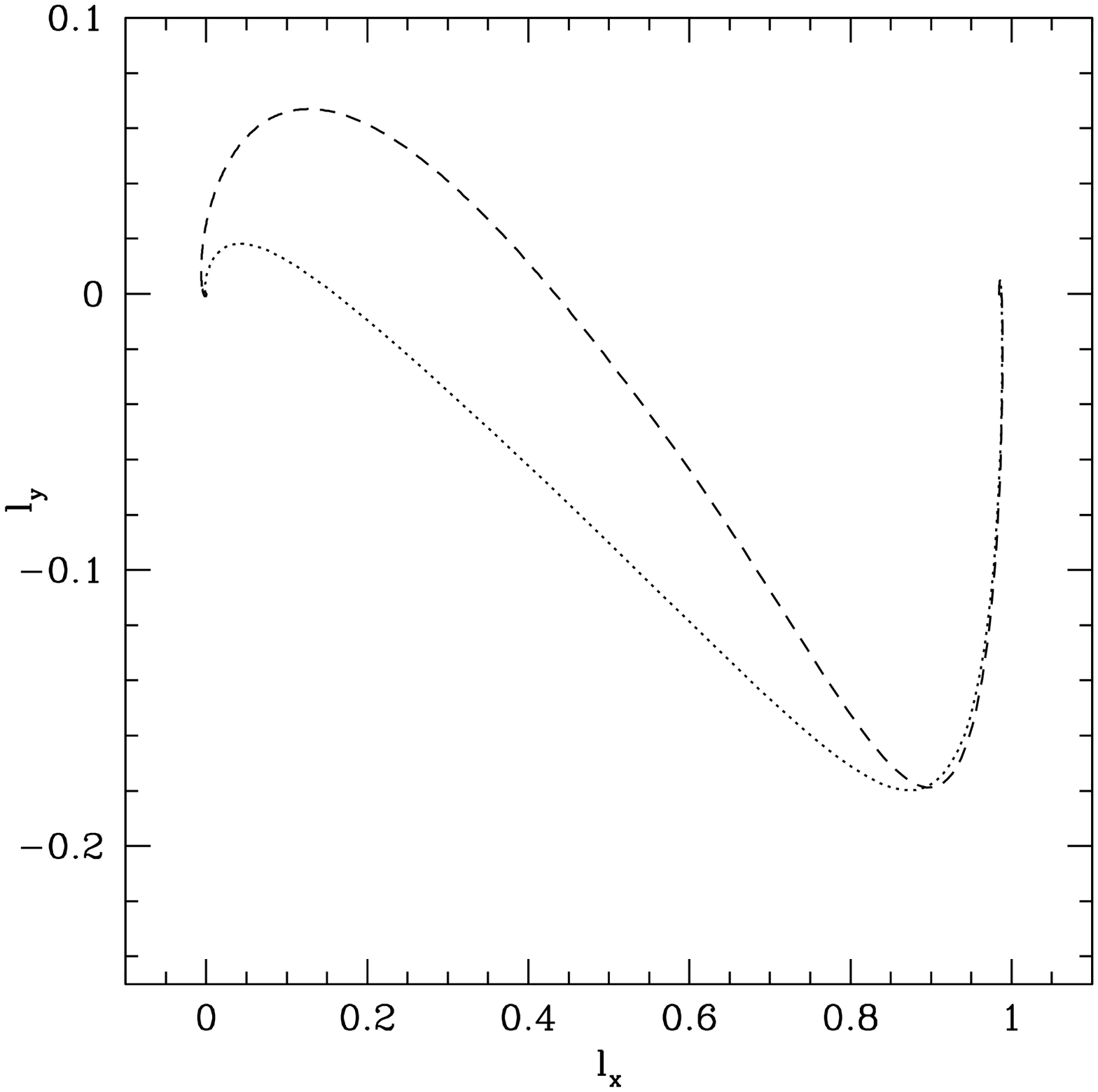}
  \caption[] {Models with the black hole 
    spin misaligned to the binary orbital axis by $\theta_0=80^\circ$.
    Models show an accretion disc with both the Lense-Thirring effect
    and a binary torque term with $R_{\rm LT}=10\,R_{\rm in}$, $R_{\rm
      tid}=10^2\,R_{\rm in}$ and $R_{\rm out}=10^4\,R_{\rm in}$.  The
    dotted lines are a numerical model with $\Sigma=\dot
    M/(3\pi\nu_1)$ at $R_{\rm in}$ and $R_{\rm out}$ which corresponds
    to the analytic boundary conditions. The dashed line is an
    equivalent analytical matched model. Both models have $\nu_1,
    \nu_2 \propto R^2$.}
\label{num4}
\end{figure*}

In Fig.~\ref{num4} we plot the inclination and twist of a disc which
is misaligned by $\theta_0=80^\circ$. We use only the boundary conditions
$\Sigma=\dot M/(3\pi\nu_1)$ at $R_{\rm in}$ and $R_{\rm out}$. At this
high inclination the non-linear terms have a bigger effect and the
model with the same boundary conditions as the analytic solution is
not such a good fit.  

\section{Analytical Models }
\label{an}

To obtain a full solution to the warped disc problem it is necessary
to resort to numerical means. However it is possible to obtain a good
analytical approximation to the disc structure for $\theta \le
60^\circ$.  Here we construct such a solution and then investigate to
what extent it provides a satisfactory fit to the full solution.

The full equation governing the evolution of disc tilt is non-linear
but, as noted by Scheuer \& Feiler (1996), when the disc tilt is small
enough the non-linear term may be neglected. Once the equation has
been reduced to a linear one, analytical techniques may easily be
employed (see also Martin, Pringle \& Tout 2007).

For small tilts to the $z$-axis the local unit vector $\bm{l} =
(l_x,l_y,l_z)$ that describes the direction of the angular momentum
can be written to first order as $\bm{l} = (l_x, l_y, 1)$ with $l_x,
l_y \ll 1$. We then work in terms of the complex variable
\begin{equation}
W = l_x + i l_y.
\end{equation}
We take the scalar product of equation~(\ref{maineq}) with $\bm{l}$
and find
\begin{equation}
  0=\frac{1}{R}\frac{\partial}{\partial R}
  \left[3R\frac{\partial}{\partial R}(\nu_1 L)-\frac{3}{2}\nu_1L\right].
\label{int}
\end{equation}
This has the solution
\begin{equation}
\nu_1 L= C_{\rm acc}R^{1/2}+C_{\rm dec},
\end{equation}
where $C_{\rm acc}$ and $C_{\rm dec}$ are constants. The first term is
the accretion disc term for which $\nu_1 \Sigma=\,\rm const$ and the
second term is the decretion disc term which has $\nu_1 \Sigma \propto
R^{-\frac{1}{2}}$ \citep{P91}.

For an accretion disc we set $L=0$ at $R=0$ and
$L=(GM_1R)^{\frac{1}{2}}\Sigma$ everywhere so $C_{\rm dec}=0$. For a
decretion disc we have $C_{\rm acc}=0$ and so more generally we find
\begin{equation}
L=(GM_1R)^{\frac{1}{2}}\Sigma_0\left(\frac{R}{R_0}\right)^{-\beta-x},
\label{modL}
\end{equation}
where $x=0$ for an accretion disc and $x=1/2$ for a decretion disc.

\subsection{Solution}

Martin et al (2007) and Scheuer \& Feiler (1996) have demonstrated how
to obtain analytic solutions for the linear warp equation with a power
law density profile and a single added precession term. Here we wish
to add two precession terms, the Lense-Thirring precession at small
radii and the tidal precession at large radii. We have not been able
to find analytic solutions for this case. However, as we have seen
from the numerical results, at most radii in the disc  one of the
precession terms dominates. Thus we may obtain an approximate solution
to the disc by matching two solutions. At small radii we use the
solution of Martin et al (2007) for a disc with Lense-Thirring
precession. At large radii we obtain a solution for a disc with only
tidal precession. We then match these solutions at the intermediate
radius $R_{\rm warp}$ where the precession terms are of comparable but
small magnitude.

\subsection{Accretion Disc with Tidal Precession}
\label{62}

In steady state without the non-linear term and with only the
tidal torque term equation~(\ref{maineq}) becomes
\begin{equation}
  -\bm{T}_{\rm tid}
  = \frac{1}{R}\frac{\partial}{\partial R}
  \left[ \frac{1}{2}R\nu_2 L \frac{\partial \bm{l}}{\partial R}\right]
\end{equation}
and, substituting our expression for $L$ (equation~\ref{modL}) we find
\begin{equation}
  -\bm{T}_{\rm tid}
  = \frac{1}{R}\frac{\partial}{\partial R}
  \left[ \frac{1}{2} \nu_{20} 
    (GM_{\rm 1})^{\frac{1}{2}}R^{\frac{3}{2}-x}\Sigma_0 R_0^x 
    \frac{\partial \bm{l}}{\partial R}\right].
\label{balance}
\end{equation}
We add the $x$-component of equation~(\ref{balance}) to $i=\sqrt{-1}$
times the $y$-component to get $W=l_x+il_y$. Then
\begin{equation}
  -T_{\rm tid}=\frac{1}{2} \nu_{20} (GM_{\rm 1})^{\frac{1}{2}}\Sigma_0
  R_0^x 
  \frac{1}{R}\frac{d}{dR}\left[R^{\frac{3}{2}-x}\frac{dW}{dR}\right],
\label{eq}
\end{equation}
where 
\begin{equation}
T_{\rm tid}= (\bm{T}_{\rm tid})_x+i(\bm{T}_{\rm tid})_y 
\end{equation}
and the linearised tidal torque is
\begin{align}
  T_{\rm tid}= -\frac{3GM_2 \Sigma R^2}{4 R_{\rm b}^3} & \left[
    1+\frac{15}{8}\left(\frac{R}{R_{\rm b}}\right)^2 \right. \cr
  &\left. + \frac{175}{64}\left(\frac{R}{R_{\rm
          b}}\right)^4+... \right]i W
\label{tid}
\end{align}
up to $O(W^2)$ with equation~(\ref{laplace}). Here $R_{\rm b}$ is the
binary separation and we have used $l_z \approx 1$ \citep{OD01}.

We only take the first term in the power series.  This introduces an
error which is relatively large in the outer parts of the disc.
However, our outer boundary condition means that the torque itself at
the outer edge of the disc is zero. We expect the disc to remain
aligned with the binary orbit until $R \ll R_{\rm b}$ so the actual
error remains small.

With the binary torque warp radius (equation~\ref{btw}) we simplify
equation~(\ref{eq}) governing the shape of the disc to
\begin{equation}
  iW y^{2-\beta-x}=\frac{1}{y}\frac{d}{dy}
  \left[y^{\frac{3}{2}-x}\frac{dW}{dy}\right],
\label{norm}
\end{equation}
where $y=R/R_{\rm tid}$. We solve this equation for accretion discs
with $x=0$ to find the shape of each disc.

For an accretion disc with $x=0$, equation~(\ref{norm}) becomes
\begin{equation}
iW y^{3-\beta}=\frac{d}{dy}\left[y^{\frac{3}{2}}\frac{dW}{dy}\right].
\label{norm1}
\end{equation}
We make the substitution $z=y^{\frac{7-2\beta}{4}}$ to find
\begin{equation}
  iW \left(\frac{4}{7-2\beta}\right)^2 z^{\frac{9-2\beta}{7-2\beta}}=
  \frac{d}{dz}\left( z^{\frac{9-2\beta}{7-2\beta}}\frac{dW}{dz}\right). 
\end{equation}
We then set $W=y^{-1/4}V=z^{-\frac{1}{7-2\beta}}V$ to find
\begin{equation}
  z^2 \frac{d^2V}{dz^2}+z\frac{dV}{dz}-V\left[\frac{1}{(7-2\beta)^2}
    +i\left(\frac{4}{7-2\beta}\right)^2 z^2\right]=0.
\end{equation}
This is the modified Bessel equation with solution
\begin{equation}
V=A \:I_{\frac{1}{7-2\beta}}\left( \frac{4 \sqrt{i}}{7-2\beta}z\right) + 
B \: K_{\frac{1}{7-2\beta}}\left( \frac{4 \sqrt{i}}{7-2\beta}z\right),
\end{equation}
where $I$ and $K$ are modified Bessel functions of the first and
second kind and $A$ and $B$ are, possibly complex, constants of
integration. Substituting for our original variables we find
\begin{align}
  W_{\rm tid} = & \,\, y^{-1/4}\left[ A \: I_{\frac{1}{7-2\beta}}\left( \frac{4
        \sqrt{i}}{7-2\beta}y^{\frac{7-2\beta}{4}}\right)\right. \cr &
  \left. + B \: K_{\frac{1}{7-2\beta}}\left( \frac{4
        \sqrt{i}}{7-2\beta}y^{\frac{7-2\beta}{4}}\right)\right].
\label{acceq}
\end{align}

We expect the outer edge of the disc to align with the binary orbit.
Thus we expect $W\rightarrow 0$ as $R\rightarrow R_{\rm b}$ in the
frame of the orbit.  The outer boundary condition should be $dW/dR=0$
at $R=R_{\rm out}$. However when $R_{\rm out} \gg R_{\rm tid}$ then we
can use $W=0$ at $R=R_{\rm out}$.

\subsection{Matching the solutions}
\label{sec:match}

We expect the solution $W_{\rm tid}$ to be valid at large disc
radii. At small disc radii we make use of the solution given by Martin
et al (2007) for a disc warped by Lense-Thirring precession.

The general solution in this case was found to be
\begin{align}
  \tilde{W}_{\rm LT} = & r^{-\frac{1}{4}}\left[
    C \: K_{\frac{1}{2(1+\beta)}}\left(\frac{2\sqrt{i}}{1+\beta}
      r^{-\frac{1}{2}(1+\beta)}\right) \right. \cr &\left .  + D \:
    I_{\frac{1}{2(1+\beta)}}\left(\frac{2\sqrt{i}}
      {1+\beta}r^{-\frac{1}{2}(1+\beta)}\right) \right]
\label{bp}
\end{align}
\citep{MPT07} in the frame in which the spin of the black hole is in
the $z$--direction $r=R/R_{\rm LT}$. Here again $I$ and $K$ are
modified Bessel functions of the first and second kinds and $C$ and
$D$ are, possibly complex, constants of integration.  A disc subject
to both Lense--Thirring and tidal torques is dominated by this
shape when $R \ll R_{\rm warp}$.

In order to match solutions we need first to transform $\tilde{W}_{\rm
  LT}$ to the frame of the binary orbit, so that the axes of both
solutions are aligned. Note that we still formally require all
misalignment angles to be small. To transform this solution to the frame
of the binary we obtain
\begin{equation}
  W_{\rm LT}=\cos \theta_0 \Re(\tilde{W}_{\rm LT})+
  \sin \theta_0 \sqrt{1-|\tilde{W}_{\rm LT}|^2}+i\Im(\tilde{W}_{\rm LT}),
\end{equation}
where $\theta_0$ is the angle between the spin axis of the black hole
and the normal to the orbital plane. We note that we are introducing
non-linear terms by this transformation but because it is only a
rotation of the axes it does not affect the shape.

We now take these two solutions, the inner solution $W_{\rm LT}$ and
the outer solution $W_{\rm tid}$, and match them at the intermediate
radius $R_{\rm warp}$, the radius at which the two precession terms
have an equally small effect. In this way we find a matched analytical
solution valid at all radii throughout whole disc. We are working in
the frame of the binary. We expect the disc to be aligned with the
black hole spin and so take
\begin{equation}
W_{\rm LT}({R_{\rm in}})=\sin \theta_0.
\end{equation}
This gives one relation between the constants of integration $C$
and $D$. At the outer edge we expect the disc to be aligned with the
binary orbit and so take
\begin{equation}
W_{\rm tid}({R_{\rm out}})=0.
\end{equation}
This gives one relation between the constants of integration $A$
and $B$.

At $R=R_{\rm warp}$ we require the solution for $W$ to be continuous
so that
\begin{equation}
W_{\rm LT}(R_{\rm warp})=W_{\rm tid}(R_{\rm warp}).
\end{equation}
Because $W$ satisfies a second order differential equation we also
require that the derivative of $W$ be continuous so that
\begin{equation}
\frac{dW_{\rm  LT}}{dR}(R_{\rm warp})=\frac{dW_{\rm tid}}{dR}(R_{\rm warp}).
\end{equation}
These two conditions give two further relations between the constants
$A$, $B$, $C$ and $D$ so we now have enough to solve for the complex
constants $A$, $B$, $C$ and $D$ and find the complete matched solution
for the disc. The form of the constants is complicated and does not
add anything particular to our understanding so we don't reproduce
them here. We do however construct many such matched solutions and
compare them with our numerical models.

\section{Accuracy of the analytic solutions}

We plot these analytical solution in the dashed lines for the
parameters of the numerical models in Fig.~\ref{num}, \ref{num2},
\ref{num3} and~\ref{num4} alongside the numerical calculations. At an
inclination $\theta_0=20^\circ$ the solutions match very well. Even at
$\theta=60^\circ$ they match well when the same boundary conditions
are used. Though deviations are larger at $80^\circ$ much of the shape
is still reproduced. We note that the analytical solutions are easily
found for any $\beta$. The numerical solution takes much longer to
evaluate when $\beta$ is small and the timestep, controlled by the
centre of the disc, becomes very small.

\subsection{Error in neglecting the non-linear terms}

\begin{figure}
 \epsfxsize=8.4cm \epsfbox{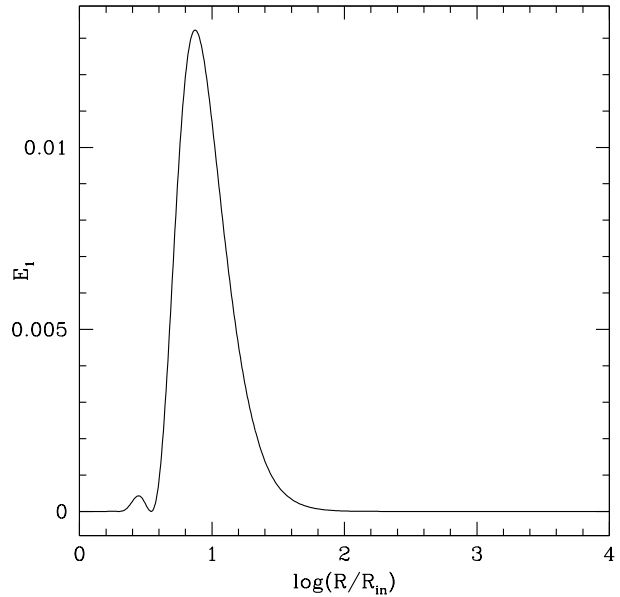}
  \caption[] {The effect of neglecting the non-linear 
    terms in the matched analytical solution for a disc with the spin
    axis of the black hole misaligned with the binary orbital axis by
    $\theta_0=20^\circ$. We choose $R_{\rm LT}=10\,R_{\rm in}$,
    $R_{\rm tid}=100\,R_{\rm in}$ and $\nu_{2}/\nu_1=10$ as in
    Fig.~\ref{num}. }
\label{matcherror}
\end{figure}

\begin{figure}
 \epsfxsize=8.4cm \epsfbox{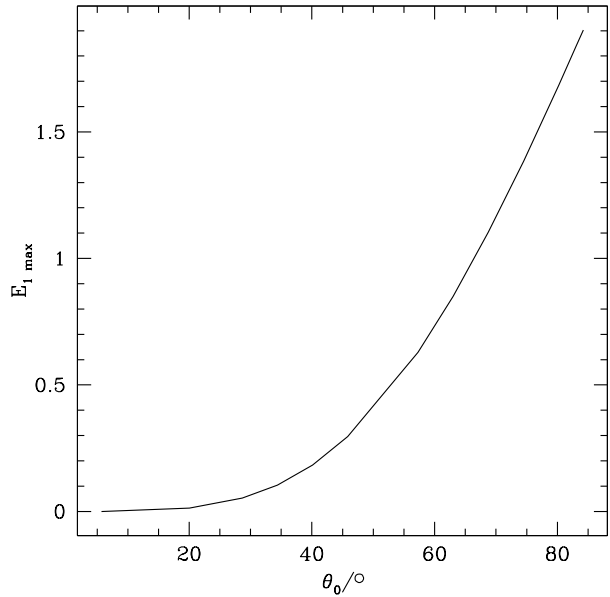}
  \caption[] {The maximum error in neglecting the non-linear terms 
    for a model with $R_{\rm LT}=10\,R_{\rm in}$ and $R_{\rm
      tid}=100\,R_{\rm in}$ and with $\nu_{2}/\nu_1=10$ for a range of
    misalignments $\theta_0$ between the black hole spin and the
    orbital angular momentum of the binary system. }
\label{peak}
\end{figure}

We have assumed that the warp is gradual enough that we can neglect
the non-linear term $\bm{l}.\partial^2\bm{l}/\partial R^2=
-\left|\partial \bm{l} / \partial R \right|^2$. The relative error in
neglecting this term is given by
\begin{equation}
E_1=\left|\frac{\frac{1}{2} \nu_2 L \frac{\partial^2 \bm{l}}{\partial R^2}\bm{.l}}
{\frac{3}{2}\frac{1}{R}\frac{\partial}{\partial R}(\nu_1 \bm{L})\bm{.l}}\right|.
\label{e1}
\end{equation}
We can justify neglecting this term if $E_1\ll 1$. With similar
power-law viscosities 
\begin{equation}
  E_1= \frac{2}{3}\left(\frac{\nu_{2}}{\nu_{1}}\right) R^2 
  \left| \frac{\partial \bm{l}}{\partial R}\right|^2.
\end{equation}
We plot this for the matched solution for a misalignment of
$\theta_0=20^\circ$ in Fig.~\ref{matcherror}. We see that the error is
at a maximum of just over $1\,$per cent at $R=R_{\rm LT}$. In
Fig.~\ref{peak} we plot the maximum value of $E_1$ for different
misalignment angles between the black hole spin and the binary orbit.
It begins to dominate after $\theta_0=60^\circ$ and this accounts when
most of the error in the model for $\theta=80^\circ$.

\section{Shape of the disc}

\begin{figure}
 \epsfxsize=8.4cm \epsfbox{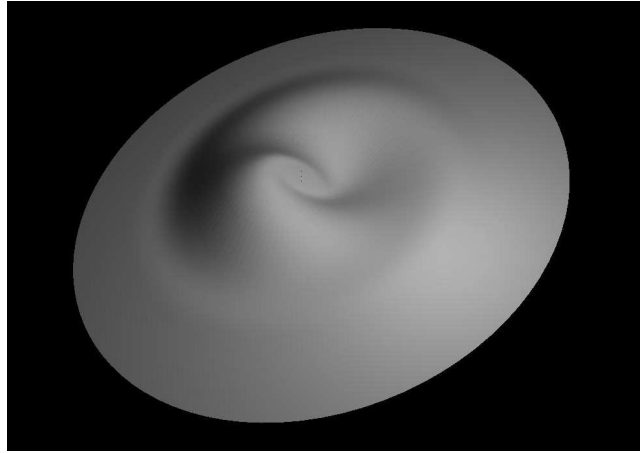}
  \caption[] {A picture of the shape of a disc with the 
    matched analytical disc solutions with $R_{\rm LT}=10\,R_{\rm
      in}$, $R_{\rm tid}=100\,R_{\rm in}$ and a misalignment angle of
    $\theta_0=45^\circ$. The radial scale is logarithmic. }
\label{pic}
\end{figure}

Given a position in cylindrical polar coordinates $(R,\phi_i)$ in the
$x-y$ plane of the binary orbit, we can find the height, $z$, of the
disc at that point. In Cartesian coordinates the disc can be described
by
\begin{align}
x & =R \sin \theta \cos \phi_i \cr
y & =R \sin \theta \sin \phi_i \cr
z & =R \cos \theta \cos(\phi_i-\phi),
\end{align}
where $\theta(R)=\cos^{-1} (l_z)$ is the angle of the direction vector
of the disc, $\bm{l}$, at radius $R$, to the $z$-axis. The azimuthal
angle of $\bm{l}$ is $\phi=\tan^{-1} (l_y/l_x)$.  In Fig.~\ref{pic} we
picture this disc solution, with a logarithmic scale in the radial
direction and $R_{\rm in}$ mapped to the origin, using the plotting
package {\sc Povray}. We can see the warp and twist of the disc for a
disc with a misalignment of $\theta_0=45^\circ$ with the warp radii
$R_{\rm LT}=10\,R_{\rm in}$ and $R_{\rm tid}=100\,R_{\rm in}$.

\begin{figure*}
 \epsfxsize=8.4cm \epsfbox{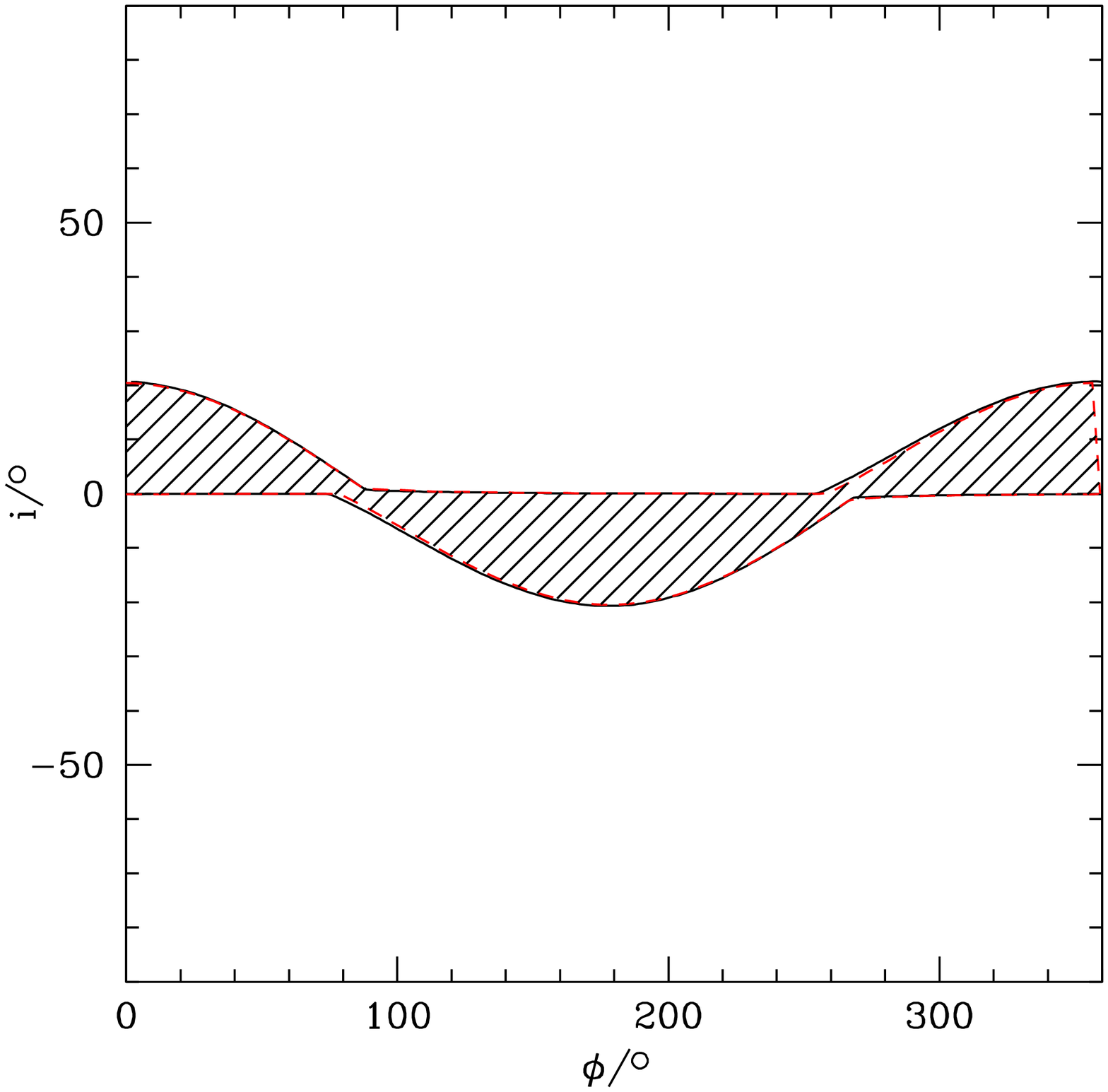}
 \epsfxsize=8.4cm \epsfbox{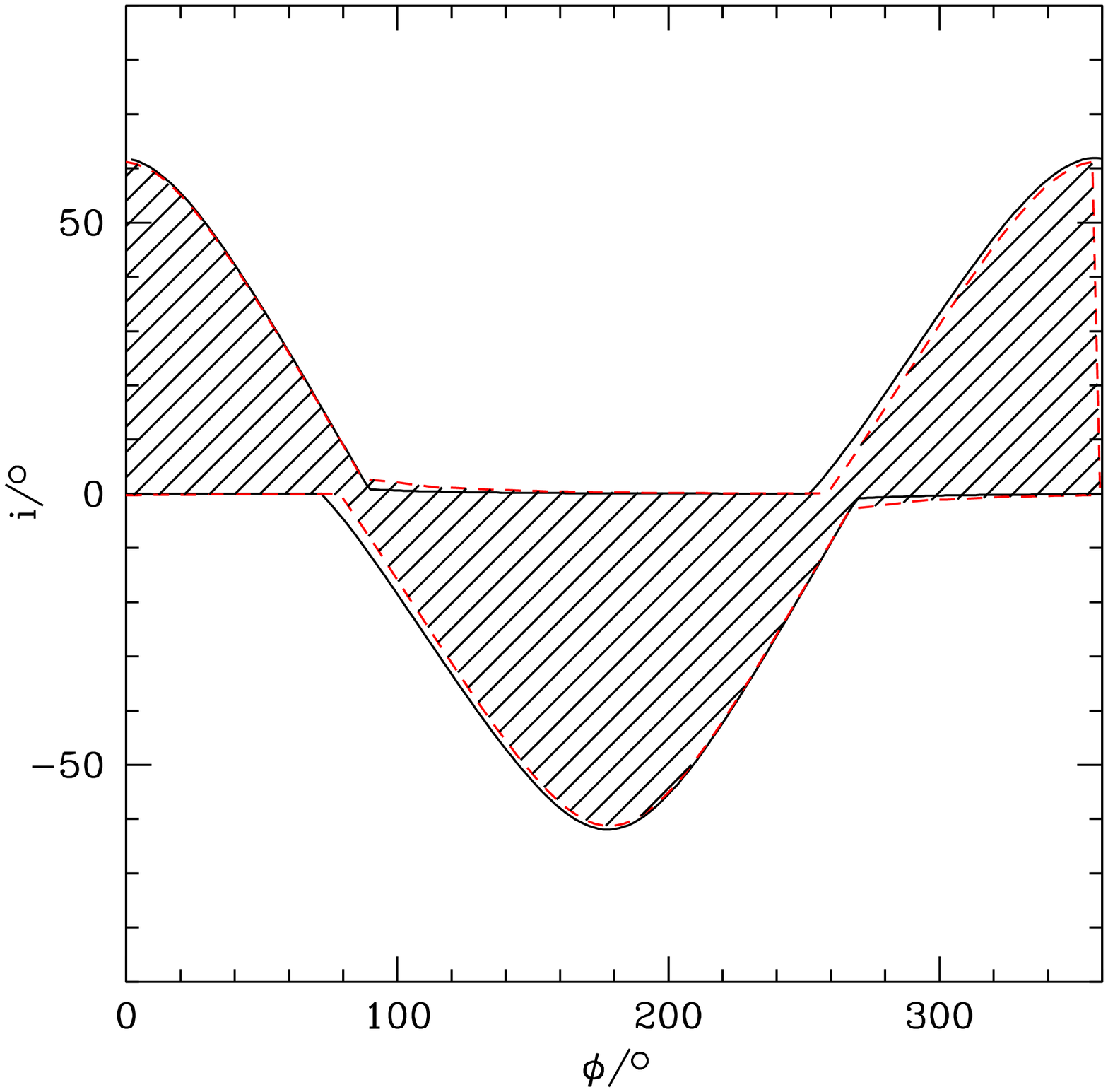}
  \caption[] {The visibility of the black hole with a disc with the 
    numerical (solid lines) and matched analytical disc solutions
    (dashed lines) with $R_{\rm LT}=10\,R_{\rm in}$, $R_{\rm
      tid}=100\,R_{\rm in}$ and a misalignment angle of
    $\theta_0=20^\circ$ (left) and $60^\circ$ (right). The binary
    orbit is inclined at $i$ to the line of sight and $\phi$ is the
    phase around the disc. The black hole itself and the central parts
    of the disc are obscured between the two lines (the shaded
    region). }
\label{height}
\end{figure*}

These models can be used to investigate disc properties and compare
with observations. We illustrate this by examining the obscuration of
the central black hole by the disc at various inclinations.  In
Fig.~\ref{height} we plot the visibility of the central black hole on
the sky for $\theta_0=20^\circ$ and $60^\circ$ for the numerical model
with zero torque boundary conditions and for the analytically matched
solution. The binary orbital axis is inclined at $i$ to the line of
sight. The angle $\phi$ is the phase around the disc. It changes on
the precession timescale of the black hole \citep{MTP08}.  This is
long compared with the orbital period so it is essentially fixed
relative to the observer. As expected, the higher the misalignment,
the more obscured the black hole is. In each plot the region between
the two lines would be obscured. Known systems would therefore tend to
be observationally selected to be aligned.

\section{Conclusions}

We have made both numerical and analytical models of accretion discs
warped by the Lense-Thirring effect and a binary companion. The models
have $\nu\propto R^\beta$ with $\beta=2$ so that the viscous timescale
is constant throughout the disc. This means that the models converge
at their fastest rate. More realistically $\beta=3/4$ and the
numerical disc evolves slowly because the small viscous timescale at
the inner edge dominates. On the other hand our analytical matched
models can be constructed just as quickly for any $\beta$. Up to a
misalignment of about $\theta=60^\circ$ they provide a very good fit
to the disc structure and so may be used where many calculations are
required in a short time. This is necessary when attempting a fit to
observations with unknown misalignment. With a relatively simple
analytical fit it is easy to compute all sorts of disc properties such
as the extent of irradiation by the central engine at any point or the
integrated spectrum without the need to compute a numerical model for
every set of parameters. We have found that known systems are
observationally selected to be aligned.

\section*{Acknowledgements}
We thank the referee for carefully reading the manuscript and useful
comments. CAT thanks Churchill College for a Fellowship.

\label{lastpage}

\begin{thebibliography}{99}

\bibitem[\protect\citeauthoryear{Bailes}{1989}]{B89}
Bailes M., 1989, ApJ, 341, 917
        
\bibitem[\protect\citeauthoryear{Bardeen \& Petterson}{1975}]{BP}
  Bardeen J. M., Petterson J. A., 1975, ApJ, 195, L65

\bibitem[\protect\citeauthoryear{Brandt \& Podsiadlowski}{1995}]{BP95}
Brandt N., Podsiadlowski P., 1995, MNRAS, 274, 461

\bibitem[\protect\citeauthoryear{Dewey \& Cordes}{1987}]{DC87}
Dewey R. J., Cordes J. M., 1987, ApJ, 321, 780

\bibitem[\protect\citeauthoryear{Greene, Bailyn \& Orosz}{2001}]{G01}
  Greene J., Bailyn C. D., Orosz J. A., 2001, ApJ, 554, 1290

\bibitem[\protect\citeauthoryear{Hjellming \& Rupen}{1995}]{H95}
Hjellming R. M., Rupen M. P., 1995, Nat, 375, 464


\bibitem[\protect\citeauthoryear{King et al.}{2005}]{KLOP} King A. R.,
  Lubow S. H., Ogilvie G. I., Pringle J. E., 2005, MNRAS, 363, 49

\bibitem[\protect\citeauthoryear{Kumar \& Pringle}{1985}]{KP85} Kumar
  S., Pringle J. E., 1985, MNRAS, 213, 435

\bibitem[\protect\citeauthoryear{Lodato \& Pringle}{2007}]{LP07}
Lodato G., Pringle J. E., 2007, 381, 1287

\bibitem[\protect\citeauthoryear{Lubow \& Ogilvie}{2000}]{L00}
Lubow S. H., Ogilvie G. I., 2000, ApJ, 538, 326

\bibitem[\protect\citeauthoryear{Martin, Pringle \&
    Tout}{2007}]{MPT07} Martin R. G., Pringle J. E., Tout C. A., 2007,
  MNRAS, 381, 1617

\bibitem[\protect\citeauthoryear{Martin, Tout \& Pringle}{2008}]{MTP08}
Martin R. G., Tout C. A., Pringle J. E., 2008, MNRAS, 387, 188

\bibitem[\protect\citeauthoryear{Martin, Reis \&
    Pringle}{2008}]{MRP08} Martin R. G., Reis R. C., Pringle J. E.,
  2008, MNRAS, 319, L15

\bibitem[\protect\citeauthoryear{Martin, Tout \& Pringle}{2009}]{MTP09}
Martin R. G., Tout C. A., Pringle J. E., 2009, in press

\bibitem[\protect\citeauthoryear{Mirabel \& Rodr\'{i}guez}{1999}]{MR99}
Mirabel I. F., Rodr\'{i}guez L. F., 1999, ARA\&A, 37, 409

\bibitem[\protect\citeauthoryear{Ogilvie \& Dubus}{2001}]{OD01}
Ogilvie G. I., Dubus G., 2001, MNRAS, 320, 485

\bibitem[\protect\citeauthoryear{Ogilvie}{1999}]{Ogilvie99}
Ogilvie G. I.,  1999, MNRAS, 304, 557

\bibitem[\protect\citeauthoryear{Ogilvie}{2000}]{Ogilvie00}
Ogilvie G. I.,  2000, MNRAS, 317, 607

\bibitem[\protect\citeauthoryear{Orosz et al.}{2001}]{O01} Orosz
  J.~A., Kuulkers E., van der Klis M., McClintock J.~E., Garcia
  M.~R.,Callanan P.~J., Bailyn C.~D., Jain R.~K., Remillard R.~A.,
  2001, ApJ, 555, 489

\bibitem[\protect\citeauthoryear{Papaloizou \& Pringle}{1983}]{PP83}
Papaloizou J. C. B., Pringle J. E., 1983, MNRAS, 202, 1181

\bibitem[\protect\citeauthoryear{Pringle}{1991}]{P91} 
Pringle J. E., 1991, MNRAS, 248, 754

\bibitem[\protect\citeauthoryear{Pringle}{1992}]{P92} 
Pringle J. E., 1992, MNRAS, 258, 811

\bibitem[\protect\citeauthoryear{Scheuer \& Feiler}{1996}]{SF}
Scheuer P. A. G., Feiler R., 1996, MNRAS, 282, 291

\bibitem[\protect\citeauthoryear{Shklovskii}{1970}]{S70}
Shklovskii I. S., 1970, SvA, 13, 562

\bibitem[\protect\citeauthoryear{Sutantyo}{1978}]{S78}
Sutantyo W., 1978, Ap\&SS, 54, 479

\end{thebibliography}
\end{document}